\documentclass[lettersize,journal]{IEEEtran}
\usepackage{amsmath,amsfonts}
\usepackage{algorithmic}
\usepackage{algorithm}
\usepackage{array}
\usepackage[caption=false,font=normalsize,labelfont=sf,textfont=sf]{subfig}
\usepackage{textcomp}
\usepackage{stfloats}
\usepackage{url}
\usepackage{verbatim}
\usepackage{graphicx}
\usepackage{cite}
\usepackage{xcolor}
\begin{document}

\title{5/6G: Networks of the Future \\or Defuturing Networks?}

\author{Cristina Cano and Hug March
\thanks{C. Cano is with the Estudis d'Informàtica Multimèdia i Telecomunicació \& Internet Interdisciplinary Institute (IN3), UOC. H. March is with the Estudis d'Economia i Empresa \& Internet Interdisciplinary Institute (IN3), UOC
}
\thanks{Manuscript received XXX; revised XXX.}}

\markboth{Journal of \LaTeX\ Class Files,~Vol.~XX, No.~XX, XX~XXXX}%
{Shell \MakeLowercase{\textit{et al.}}: A Sample Article Using IEEEtran.cls for IEEE Journals}

\IEEEpubid{}

\maketitle

\begin{abstract}
 
Energy efficiency is at the core of sustainability solutions for 5/6G networks. We argue this is a too narrow perspective on sustainability, as it ignores the effects of the increased traffic demand these networks stimulate and the need for additional equipment that this demand requires. The hope is that techniques to reduce the network's energy consumption in operation will be able to compensate for increases in traffic demand. However, we argue that there are more challenges than just reducing the energy that the network requires to function and that it is not clear whether higher energy efficiency will be able to cope with increasing demand. The need for more equipment related to deployment of 5/6G networks (both at the user end, network level and the cloud and grid) may result in important environmental impacts related to: i) increased pressures on material extraction, which imply socio-environmental conflicts, ecosystem destruction and displacement, ii) more manufacturing and shipment, with effects on greenhouse gas emissions and pollution; iii) increased disposal complexities and challenges to recycle components of such equipments. By extending our view on sustainability to include the aforementioned often ignored implications, we are able to identify design requirements and technical pillars of 5/6G networks  that need to be rethought. We also devise new paths forward to address these challenges. We argue that it is crucial to think of alternative applications and requirements that aim to serve user demands explicitly, instead of incessantly creating new needs. We also claim that acknowledging material limitations in the production of new hardware is critical, promoting retrofitting and modular design in future network development. The conclusions of this article show that it is time to start rethinking the course of mobile network development in order to align it to current environmental objectives to tackle the climate emergency the world is experiencing.    
 
\end{abstract}



\section{Introduction}
\IEEEPARstart{E}{nergy} consumption is the key axis structuring the narratives of the industry and policy makers when thinking about the relationship between ICT and sustainability, see \cite{vhk2020ict}. However, ICT has far-reaching environmental, social and economic consequences not limited to the energy consumption during its lifecycle. 

It is well-known that rare earth materials are needed for the production of ICT equipment. While these \emph{rare} elements are present in many geographies across the globe, the political geography of their processing is complex, and their extraction and processing may cause important socio-environmental problems. These include socio-environmental conflicts, dubious labour practices, environmental destruction and displacement \cite{crawford2021atlas}. The case of the Fairphone shows that replacing current used elements with sustainable counterparts is hard, since, although they are committed to use sustainable materials, these only account for 15\% of the product \cite{fors2019problematizing}. On top of that, the extraction and processing of resources essential for key components of ICT, such as batteries (e.g. lithium), should also be taken into account. Beyond the extraction of some critical materials for ICT, the manufacturing of ICT equipment is another source of socio-environmental problems. Studies for specific equipment have shown that the production stage accounts for an important part of the energy consumption of the lifecycle of ICT products \cite{fors2019problematizing}. Additionally, the shipment of equipment is usually neglected when analysing the sustainability of ICT. However, shipment is responsible for 3.1\% of global carbon dioxide emissions and impacts on the oceanic ecosystem as thousands of containers are lost each year, sometimes releasing toxic substances into the water \cite{crawford2021atlas}. What to do when equipment is no longer needed, or it is replaced, is another significant concern of ICT in terms of sustainability. The complexity of current equipment makes it hard to recycle. Because of this, most of this equipment is sent to the Global South to be used as refurbished goods (which are informally recycled or disposed of in landfills after use) or for informal recycling, with strong effects on the environment, the workers and the local communities \cite{fors2019problematizing}.

When we consider mobile network technologies specifically, we observe that there has been renewed interest in energy efficiency, especially in 6G, where energy efficiency has become a design requirement \cite{chowdhury20206g}. The focus is on reducing the network’s energy consumption in operation by designing energy-efficient techniques that allow for a reduced energy use per amount of information transmitted. We argue that this perspective is a too narrow stance on sustainability, as it ignores indirect as well as embodied energy sources \cite{williams2022energy}. On the one hand, it is not clear that energy efficient techniques will be able to cope with the increasing traffic demand that these networks stimulate. On the other hand, this vision neglects the increasing need for equipment in terms of impact on greenhouse gas emissions, socio-environmental conflicts, pollution and destruction of ecosystems we overviewed above. To have a sense of the increased need for equipment, consider Dell’Oro Group Telecom Capex Report, which shows that telco Capex (capital expenditures) have increased 9\% in 2021 (reaching more than US\$60 billions) and is expected to increase 3\% in 2022 before gradually becoming smaller in 2023 \cite{delloro2022report}, presumably after the initial roll-out of 5G network equipment.  

In this article, we scrutinize 5/6G network design through this broader sustainability perspective and question whether these networks that start shaping the present and will be the future of mobile communications are, in fact, defuturing. That is, whether they contribute to the destruction of the future by design \cite{fry2020defuturing}, encouraging unlimited consumerism irrespective of the planetary boundaries we are about to surpass, including climate change  and the aim to limit global warming to 1.5/2ºC. By taking this expanded vision on sustainability, we are able to identify specific design requirements and technical innovations of 5/6G networks that are of concern and devise alternative socio-environmental ways forward not guided by the logic of endless economic growth. 

Even though the environmental benefits of 5/6G networks are not yet clear \cite{williams2022energy}, these networks, as well as other ICT innovations, are increasingly viewed not as a sustainability problem that needs to be addressed, but as the solution to the current climate crisis \cite{fors2019problematizing} due to the effects they can have on other domains (e.g. bringing more efficiency in terms of production and consumption of other goods and services). The conclusions we draw in this article suggest that it is time to start reconsidering the position of researchers and developers towards network development based on incentivizing ever-increasing and stringent demand in a range of sectors driven by endless replacement of equipment and production of new devices. The broader sustainability perspective on the implications of these networks we present in this article raises doubts on whether the energy savings they may enable in other sectors would be able to compensate for the environmental negative tradeoffs exposed above. 

The rest of the article is organized as follows. In Section II, we overview the different applications and technical pillars of 5/6G communication networks. Then, in Section III, we analyse the implications of the design of these networks under the broader sustainability perspective we take in this article. After this, in Section IV, we devise alternative ways forward that take account of this wider sustainability perspective. Finally, we provide some final remarks. 

\section{5/6G Networks}

In this section, we overview the different applications that 5/6G networks aim to provide and the technical pillars on which the new services are based.

\subsection{Applications}

The main objective of 5/6G networks is to provide a myriad of applications with ever-increasing capacity, latency and data processing demands. While 5G was designed to provide applications such as two-way gaming, tactile Internet, virtual reality and autonomous driving \cite{andrews2014will}, 6G goes even further, including applications that use brain-computer interfaces, implants, smart wearables, haptics, extended reality and flying vehicles \cite{chowdhury20206g}.

These applications imply that the network must be able to provide increased capacity, reduced latency and the management of heterogeneous services. Nevertheless, they also imply a massive increase of end devices as well as the need to handle, from the network side, the connections and data they generate.

\subsection{Technical pillars}

In order to provide the capacity and latency demands of the envisioned applications and manage the different services and increased number of devices, 5/6G networks are based on several technical pillars. We overview next what we consider the most important ones to deliver the exposed requirements:

\subsubsection{Higher frequencies}

One of the key pillars to providing increased capacity in 5/6G networks is the use of higher frequencies. In 5G the approach is to use mmwave bands (in the order of GHz) while in 6G the plan is to use even higher frequencies (THz bands, above 300 GHz \cite{chowdhury20206g}). 

\subsubsection{More antennas}

Another important design for providing increased capacity is the use of a higher number of antennas. In this case, the innovation in 5G was the use of massive MIMO, while in 6G this is extended to large intelligent surfaces \cite{chowdhury20206g}, in which antenna arrays are integrated into large structures.

\subsubsection{Network slicing}

In order to handle the myriad of heterogeneous services, 5G networks are implementing network slicing, enabled by software-defined networking and network function virtualization. This cloud computing paradigm in network management allows dedicated service delivery and the provision of heterogeneous services with various levels of requirements \cite{chowdhury20206g}.

\subsubsection{Artificial intelligence}

The orchestration of network management relies on artificial intelligence (AI) in 5/6G networks in order to dynamically and optimally allocate network resources in a centralized manner. While 5G networks are said to implement limited AI, 6G networks are expected to support full AI automation \cite{chowdhury20206g}.

In summary, the combination of higher frequencies and an increasing number of antennas, along with a more flexible and optimized management of the network, are the key elements of 5/6G networks to enable the provision of heterogeneous and demanding services. 

6G incorporates other innovations which are expected to also play a role in providing high-demanding services. The need to increase the capacity of the backhaul in order to cope with increased service demand is addressed in 6G with the use of optical wireless and backhaul communications \cite{chowdhury20206g}. Intelligent reflecting surfaces are also a key innovation in 6G \cite{chowdhury20206g}, which changes the paradigm we have seen until now. The focus of previous generations was to improve the transmitter and receiver ends, considering the environment alien to the design of the network. Intelligent reflecting surfaces allow modifying the propagation conditions in order to optimize network performance using either passive or active reflecting elements to our surroundings. 3D networking is also a new research direction in 6G \cite{chowdhury20206g}. Under this approach, new communication elements are added to the network at different altitude levels, including drones and satellite devices. The integration of transfer of data and sensing is also another innovation that is drawing attention in 6G given the increased frequency bands planned to operate in this network generation \cite{chowdhury20206g}. Finally, we can also highlight the use of blockchain and quantum communications, which are expected to be integrated into 6G \cite{chowdhury20206g}. 

\section{Are 5/6G networks defuturing?}\label{sec:defuturing}

Energy efficiency has been considered the key aspect of sustainability in cellular networking. One of the goals of 6G is to reach a ten-fold increase in energy efficiency \cite{chowdhury20206g}. However, other important aspects have been left out of the discussion. Thus, we wonder whether neglecting other implications make these networks, in fact, defuturing. That is, whether they contribute to the destruction of the future by design, as defined in [16]. When we act to defuture, according to \cite{fry2020defuturing}, is because “we have little comprehension of the complexity, ongoing consequences and transformative nature of our impacts”. In this section, our aim is to propose “a way of a learning to see and think the familiar differently” \cite{fry2020defuturing}, broadening the current perspective on sustainability including aspects not usually accounted for, such as the effects of stimulating greater demand and the increased need for equipment these networks require. Our conclusions imply that the design of 5/6G networks is not aligned with current objectives to limit climate change and address socio-environmental problems.

\subsection{Higher demand}

Induction is defined as the effect of an ICT application in stimulating increased use of a product or service \cite{ropke2010managing}. 5/6G networks are designed to encourage and support a massive number of devices and increasing use of ICT applications, putting a higher pressure on the user-side, the network, data centres and the grid \cite{gamez2021depredadores}. 

It is expected that efforts to reduce the energy consumption of the network in operation will be able to slash energy needs. However, there are two main sources of concern under an ever-increasing service demand scenario. On the one hand, by focusing on the energy efficiency of the network in operation, we fail to look at the bigger picture. The global energy demands that require increased use of the infrastructure also mean higher energy consumption at the user end as well as in the data centres that serve the content for the user and process the data. On the other hand, it is by no means clear that techniques to reduce operational energy use will be able to cope with increasing demand in the context of 5/6G networks \cite{williams2022energy}. Even if we optimize the energy spent per bit transmitted (that is, Joules/bit), the total energy expenditure will not decrease if the amount of information to transmit (bits) increases at a faster pace. 

Network researchers and developers rely on forecasts that predict increases in demand and in the number of devices as an argument to continue delivering and supporting increased use. However, we should start to account for the active role that the infrastructure plays in shaping these same future demands.

\subsection{The need for more equipment}

The energy associated with the manufacturing and installation of 5/6G networks, as well as that of associated networks and devices, is usually neglected \cite{williams2022energy}. The energy needs, geopolitical issues and socio-environmental conflicts, displacement and environmental destruction related to material extraction and processing in regions situated in, usually, the Global South, is commonly not accounted for. Also, manufacturing and shipment of equipment also imply pollution and environmental impacts, with contributions hard to measure and usually not considered \cite{crawford2021atlas}. We argue there is an urgent need to quantify and consider these implications to inform decisions on hardware replacement and to better guide further stimulation on the need for new equipment. Next, we overview the increasing need for equipment that results from 5/6G network development and that contributes to these socio-environmental problems.

\subsubsection{End devices}

To get access to the new services brought about by 5G (and in the future 6G) users need to acquire smartphones (and replace the older ones) ready for that technology. In terms of environmental impact, the manufacturing phase of smartphones dominates their carbon footprint \cite{williams2022energy}, and therefore, stimulating the rate at which smartphones are replaced needs to be reassessed. On top of that, 5/6G networks not only imply the need of smartphones ready to get the most of these new networks, but the services that 5/6G networks enable require a myriad of end devices, such as virtual and extended reality glasses, gaming gadgets, brain-computer interfaces, wearables, haptics extensions, etc. Plus, we need to account for the accessories that go along with these devices, such as, for instance, headphones.  The impact of the extraction of materials, production and shipment of new smartphones and the rest of the aforementioned devices at a global scale needs to be accounted for to have a real understanding of the sustainability of these networks and the services they enable.

\subsubsection{Network devices}

The energy consumption implied in the manufacturing of a base station has been shown to account for 10-36\% of its total energy consumption over a 10-year lifetime \cite{williams2022energy}. On the other hand, we also have to take into account the materials used during the manufacturing of the product. As we have described in the previous section, a key enabler of the applications envisioned in 5/6G networks is the use of higher frequency bands. Moving higher in frequency implies shorter links as the propagation of the signal is impaired. In these conditions, densification is required. That is, the use of more base stations to cover the same area, which implies a higher need for equipment at the network side.

In addition to more base stations, 5/6G networks also make use of a higher number of antennas, as we also overviewed in the previous section. The use of a higher number of antennas is intensified in 6G with large intelligent surfaces and intelligent reflecting surfaces. The first include an array of antennas for transmission/reception, while the latter reconfigure the space to modify propagation conditions. Again, the manufacturing impacts of these surfaces must be considered in terms of sustainability. Optical devices and 3D networking using satellites are also sources of increased need for equipment in 6G that needs to be accounted for, including all the extraction and/or transformation of all materials used in the manufacturing process. 

\subsubsection{Processing equipment}

The provision of heterogeneous and demanding services requires a flexible and dynamic architecture. As we have seen in the previous section, this orchestration relies on AI, and we expect 6G networks to fully optimize network operation. This optimization aims to make efficient use of the available resources while providing the expected capacity and latency as new services demand access. AI mechanisms rely on data and computation to predict and provide the best configuration. This implies high amounts of resource consumption used in computing infrastructures and increased needs for processing equipment.

Also, the need for edge computing to handle latency-sensitive services in 5/6G networks and the integration of blockchain and quantum computing in 6G are expected to drive further demand for the manufacturing of processing hardware, with the ensuing implications in terms of energy used and materials. 

Finally, expanding the view on the global impact, we need to account as well for the processing required to serve the user content, which implies increased manufacturing of processing units to serve the resulting demand in the cloud.

\subsubsection{Renewable energy equipment}

Relying on renewable energy does not come at zero cost when considering the extraction of materials and their manufacturing. Further reliance on renewables to improve network sustainability implies a higher need for renewable energy equipment that cannot be left out of the analysis.

In summary, when considering the sustainability of 5/6G networks, there are more concerns than just energy consumption of the network in operation, which captures much of the attention by researchers and developers. In the sustainability narratives of 5/6G networks, the effects of stimulating service demands and the need for more equipment are aspects that have been traditionally neglected. In order to account for these, we argue that there is a need for alternative views of future mobile network development that are aware of these (neglected) environmental implications. 

\section{Is there a sustainable way forward?}\label{sec:wayforward}

There is a strong focus on energy efficiency in mobile network design, especially in 6G \cite{wu2017overview}, \cite{huang2019survey} However, the effects of increased demand that these networks stimulate and the need for equipment required to function are usually not considered neither in the analysis of their sustainability nor in new sustainable proposals. In this section, we devise alternative ways forward that take into account the broader vision towards 5/6G network sustainability that we have discussed in the previous section.

\subsection{Alternative requirements and applications}

Mobile network researchers and practitioners assume, as well as stimulate, an ever-increasing need for capacity, latency and number of devices. The hopes are on energy-efficient algorithms to reduce the energy requirements to serve that demand. But, as we have seen, sustainability concerns should not be reduced to the energy use of the network in operation, but instead should have a much wider perspective, taking into account both the energy and materials throughout the manufacturing and use (and disposal) of these networks as well as those of the devices and new uses that those developments open up. 

We argue that a sustainable network approach should take this broader perspective at the very core of the services being offered in the first place. We wonder how would future networks look like if we stop focusing on delivering more capacity, connecting more devices and gathering and processing more data. The applications researchers and developers imagine not only shape the networks designed, but also drive future use and demand. 

We believe that the only way forward is to imagine other requirements and applications that are conscious of the effects of the global socio-environmental and geopolitical impacts of enabling a certain way of relating to communication technology. One way is to start envisioning how to match customer needs precisely, instead of continuously exceeding them creating new needs \cite{dabadie2021your}.

\subsection{To  stop creative destruction in network technology}

New mobile network technology has been deployed every ten years starting from the 1980s. Each generation has devalued and made obsolete the previous, which is eventually dismantled. This process of technological replacement could be understood through the schumpeterian notion of creative destruction \cite{schumpeter2010capitalism}, a critical feature of capital reproduction in late capitalism. The effects of this in network technology implies the disposal of equipment which, although being operative, is no longer of use as hardware design for different communication systems differ. The results of this incessant destruction of wealth are not equally shared, putting an extra burden on ecosystems, populations of the Global South and future generations \cite{pargman2017resource}. 

The research challenge is then to develop communication technologies that do not imply  devaluing and discarding  the existing infrastructure and devices. In other words, given the planetary boundaries we are about to surpass (or that we have in fact surpassed for some dimensions), as well as geopolitical issues that go beyond the scope of this paper, it is crucial that network development integrates into the design process a scarcity vision over the resources used and hence the implications in the future availability of hardware. We are already familiar with detaching functionality from hardware. However, this is usually done to improve performance. Is it possible to reuse hardware for other functionalities instead of replacing it? This view of extending network functionality from what is already deployed calls for modular, adaptable and reconfigurable design of hardware. Under the broad sustainability perspective we have presented, future-compatible hardware becomes a potential way forward. Last, another way forward is to design networks using the same hardware for multiple purposes, such as the current trend of integrating sensing and communication technologies \cite{chowdhury20206g}. This is again considered with the goal to increase performance, but it can also be addressed from a sustainability perspective. 

\subsection{Critical Research}

This article aims to put forward a critical research agenda around 5/6G networks and sustainability. There is an urgent need to understand the material impacts of the implementation of 5/6G networks beyond the metrics of energy consumption. In other words, there is a need to understand how the unbridled growth in data usage that 5/6G networks enable is compatible with the aim to limit energy usage (and raw materials extraction) in a context of  climate emergency and other socio-environmental crises linked to crossing the different planetary boundaries. Of course the future emissions of those networks would depend on the energy mix of a given moment and place (how much electricity production depends on fossil fuels), but we think that thinking that renewable energy alone will solve the energy problem of those networks is highly problematic (both because the limits in the implementation of renewable energy as well as its socio-environmental tradeoffs). It is also necessary to consider the material foundations of the massive development of 5/6G networks, and what geographies of extraction and socio-environmental conflict will emerge. 

Mobile network development occurs in highly specialized, technology-oriented spheres. However, in order to account for the challenges ahead, a transdisciplinary stance is needed to encompass the interrelated consequences of the promotion of unlimited capacity demands and devices in different domains and at global scale.

\section{Final Remarks}

We have argued that the focus on energy efficiency in 5/6G is a too narrow perspective on the sustainability of these networks. The effects of increased demand that these networks stimulate as well as the material needs to serve that demand need to also be accounted for. By broadening the perspective on sustainable network development, we can devise alternative ways forward. Under the current climate emergency scenario and the fragile situation with several planetary boundaries being surpassed or about to be surpassed, we need to come up with alternative requirements and applications, as well as to put a stop to continuously devaluing previous generations of networks, taking a scarcity perspective on hardware and aiming for retrofitting and modular design. To address sustainable network design, taking into account the expanded view on sustainability we have presented in this article, requires social and political action, not only technological intervention, as well as a transdisciplinary stance in network design and development.

\section*{Acknowledgments}

This project has been co-financed by the Spanish Ministry of Science, Innovation and Universities through the project RF-VOLUTION (PID2021-122247OB-I00).

\bibliographystyle{IEEEtran}
\bibliography{bibliography}

\end{document}